# Mars sedimentary rock erosion rates constrained using crater counts, with applications to organic-matter preservation and to the global dust cycle

Edwin S. Kite[1,*] and David P. Mayer[1]
1. University of Chicago.
* kite@uchicago.edu

**Abstract.**
Small-crater counts on Mars light-toned sedimentary rock are often inconsistent with any isochron; these data are usually plotted then ignored. We show (using an 18-HiRISE-image, >$10^4$ crater dataset) that these non-isochron crater counts are often well-fit by a model where crater production is balanced by crater obliteration via steady exhumation. For these regions, we fit erosion rates. We infer that Mars light-toned sedimentary rocks typically erode at ~$10^2$ nm/yr, when averaged over 10 km$^2$ scales and $10^7$-$10^8$ yr timescales. Crater-based erosion-rate determination is consistent with independent techniques, but can be applied to nearly all light-toned sedimentary rocks on Mars. Erosion is swift enough that radiolysis cannot destroy complex organic matter at some locations (e.g. paleolake deposits at SW Melas), but radiolysis is a severe problem at other locations (e.g. Oxia Planum). The data suggest that the relief of the Valles Marineris mounds is currently being reduced by wind erosion, and that dust production on Mars <3 Gya greatly exceeds the modern reservoir of mobile dust.

## 1. Introduction.

Sandblasting, aeolian infilling, and wind deflation all obliterate impact craters on Mars, complicating the use of crater counts for chronology. Aeolian resurfacing is particularly confounding for dating sedimentary rocks, because these soft materials can be rapidly eroded by the wind. Yet wind erosion of sedimentary rocks is much more than a source of noise, for four reasons. (1) Rapid exhumation by wind erosion is required for near-surface preservation of ancient complex organic matter (a target for future landers). Near-surface complex organic matter on Mars is destroyed by radiation in <$10^8$ yr, so the surface must be refreshed by exhumation (Kminek & Bada 2006, Pavlov et al. 2012, Farley et al. 2014, Grotzinger 2014). (2) The pace and pattern of recent wind erosion is a sorely-needed constraint on models of terrain-influenced aeolian erosion – i.e. landscape-wind feedbacks (Kite et al. 2013a, Day et al. 2016). (3) Wind erosion is a source of dust, and the global dust reservoir will disproportionately sample fast-eroding regions. (4) Basin-scale aeolian exhumation is intrinsically interesting. Uncommon on Earth (Heermance et al. 2013), it has probably been a dominant landscape-modifying process on Mars since 3 Gya and perhaps earlier (e.g. Bridges et al. 2014, Greeley et al. 2006, Golombek et al. 2014, Farley et al. 2014). There is direct evidence for globally-distributed saltation abrasion on Mars today. However, the extent to which the deep erosion of Mars' sedimentary rocks can be explained by uniformitarian rates and processes remains unknown.

For these four reasons, we seek to constrain erosion rates Mars sedimentary rock erosion rates, averaged over the $10^7$-$10^8$ yr timescales relevant to recent topographic change and to the preservation of complex organic matter.

The only proxy for Mars wind erosion rate that is globally available is the size-frequency distribution of impact craters. Crater-formation frequency is nearly uniform across Mars' surface (Le Feuvre & Weizcorek 2008). Therefore, crater density can be compared to a model of crater production (as a function of diameter and time) to estimate age (e.g. Michael 2013). However, the best-fit crater-production function usually deviates strongly from the observed crater size-frequency distribution (CSFD) for light-toned Mars sedimentary rocks (a subset of Mars sedimentary rocks that includes the sedimentary rock mountains in Valles Marineris and Gale; Malin & Edgett 2000). For those terrains, high-resolution images show fewer small craters than anticipated from the number of large craters (e.g. Malin et al. 2007). Moreover, sedimentary rock ages inferred from small-crater frequency can be less than those of adjacent materials that are crosscut by the sedimentary rocks. These data cannot be explained by differences in rock-target strength (Dundas et al. 2010, Kite et al. 2014). These effects appear at crater sizes up to 1 km and so cannot be attributed to limited image resolution (image data are now available at 25 cm/pixel: McEwen et al. 2010). These discrepancies are usually attributed to "resurfacing," and scientists working on Mars CSFDs either fit an age to the very largest craters on sedimentary rock terrains, or avoid sedimentary rock areas entirely (Platz 2013). Although off-isochron CSFDs have been used to explore resurfacing processes for decades (e.g. Hartman 1971, Chapman & Jones 1977), the prevailing procedure is to parameterize resurfacing as one or more events, not an ongoing process (Michael 2013 resurfacing). .

The paucity of small craters relative to large craters in easily-eroded sedimentary rock terrains can be understood if we consider resurfacing not as an event but as a process (Figure 1). In this paper, we define "obliteration" as the point beyond which a crater can no longer be identified in a high-resolution optical image (i.e. HiRISE, ~25 cm/pixel). On Mars, fresh craters with simple morphologies have a depth-to-diameter ratio ~0.2 (Melosh 1989). This relationship ensures that many crater obliteration processes (Table 1) remove small craters from the landscape more readily than larger craters. For example, suppose a landscape is being steadily and uniformly being abraded at 100 nm/yr. On such a landscape, a 20m-diameter crater (initially ~4m deep) has a lifetime of 40 Myr and a 100m-diameter (initially ~20m deep) has a lifetime of 200 Myr. Extrapolating along a hypothetical, perfect crater production function from the observed density of 100m diameter craters down to 20m diameter, one would find that the observed density of 20m diameter craters on the steadily eroding landscape is less than expected from the production function by a factor of (200 Myr)/(40 Myr)=5. This correction factor is equal to the ratio of diameters for craters <3 km (for which the initial depth of the crater is ≈proportional to the initial diameter of the crater; Watters et al. 2015). Therefore, for the steady grind-down process and for craters <3 km in diameter, and approximating the CSFD in the crater-size range of interest as

$$N(>D) = kD^{-\alpha} \qquad (1)$$

(where $D$ is crater diameter), the effect of steady-state erosion is to subtract 1 from the slope-parameter $\alpha$. It may be verified (by inspection of figures with a straight edge) that many published Mars sedimentary rock CSFDs have an "off-isochron" power-law slope that follows this rule. After the fingerprints of crater-obliteration have been identified using the parameter $\alpha$, the rate of crater-obliteration can be constrained by assuming steady-state balance between crater production and destruction.

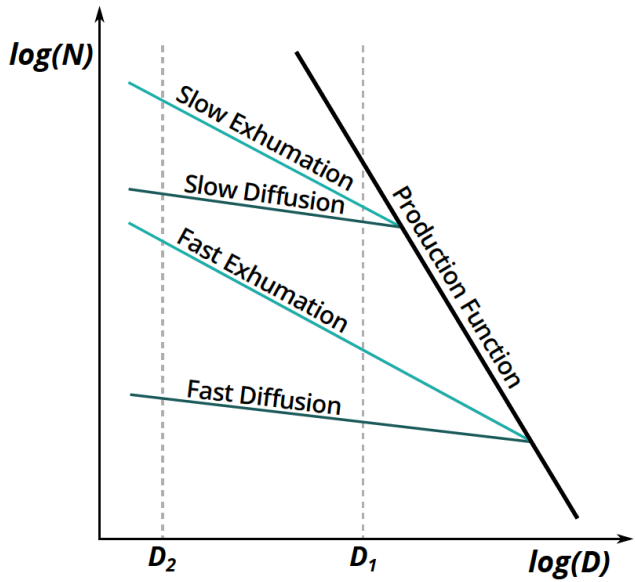

**Figure 1.** Crater frequency (N) is set by the pace of crater obliteration at a given $D$, and the slope d$N$/d$D$ is set by the process of crater obliteration. Because crater depth $d \propto D$ (crater diameter), obliteration occurs by exhumation in a time $\propto D^1$, and by diffusion in a time $\propto D^2$.

Competition between crater accumulation and obliteration has been modeled by Öpik (1965), Jones (1974), Catling et al. (2006), and Fassett & Head (2014), among others, but Smith et al. (2008) is the closest in intent to our work. Smith et al. (2008) use an analogy to radioactive decay to model size-dependent crater lifetimes for Mars craters, fitting erosion rates of ~$10^3$ nm/yr for a light-toned layered deposit at Arabia Terra and 30 nm/yr at Meridiani Planum. While we use different equations, our results are qualitatively consistent with those of Smith et al. (2008). The main differences are that we have a 100× larger crater dataset, provide a more detailed treatment of errors, consider a wider range of processes, and apply the results to a broader range of problems. Small-crater degradation has been intensively studied along the *Opportunity* traverse (Golombek et al. 2006, 2010, 2014; Fenton et al. 2015). This site is very flat, erodes slowly (3-30 nm/yr) because of armoring by hematite granules, and the CSFD is well-fit by an isochron (71±2 Myr). The *Opportunity* traverse is an outlier in that most light-toned sedimentary rocks on Mars erode quickly, are associated with steep slopes (and thus slope-winds), lack hematite armor, and have CSFDs that are not well-fit by isochrons. However, *Opportunity*'s close-up view provides constraints on small-crater degradation processes that have global relevance (Golombek et al. 2014, Watters et al. 2015): sandblasting swiftly ablates ejecta blocks and planes down crater rims, then sand-infill slowly mutes craters (left panel of Figure 2). Crater expansion during degradation is minor.

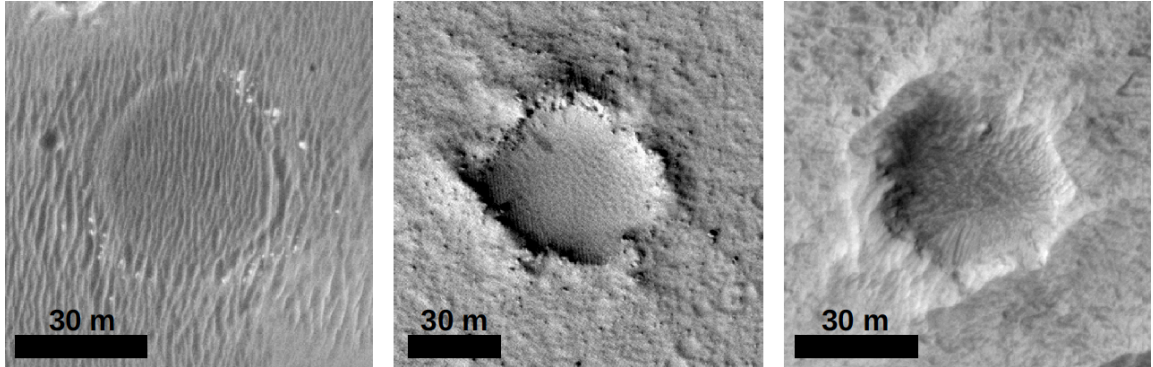

**Figure 2.** Examples of small-crater degradation style. *Left panel:* diffusive infilling."Kitty Clyde's Sister" (informal name); HiRISE image ESP_016644_1780. *Center panel:* relatively fresh crater (sharp rim remains intact around most of the crater, minimal infilling) on NW outer flank of Olympus Mons; ESP_014407_2045. *Right panel:* crater in E Candor eroded by steady exhumation; ESP_016277_1715.

This paper is about both a technique (§2-§4.1) and its application (§4.2-§8). Readers uninterested in techniques may skip to §4.2. In §2, we motivate our use of a steady-exhumation model, contrasting it with two alternatives: a one-big-pulse model and a diffusion model. Next (§3), we outline a workflow for obtaining erosion rates assuming steady exhumation. In §4, we present and analyze an example dataset obtained using 18 High Resolution Imaging Science Experiment (HiRISE; McEwen et al. 2010) images. In §5, we assess the implications of erosion rates for landscape evolution and the age of dust on Mars. In §6, we apply the resulting erosion rates to estimate organic-matter destruction. In §7, we discuss approximations, limitations, and open questions, as well as independent constraints from landslide-molds (Grindrod & Warner 2014) and cosmogenic isotopes (Farley et al. 2014). We conclude in §8.

**2. Processes and process determination.**

Fitting erosion rates to CSFDs on rocky terrain raises questions about geology (§2) and questions about methods (§3). Turning to the geology questions first:

> (1) *Are crater obliteration rates equivalent to landscape-exhumation rates?* Fresh craters have steep walls. Steep slopes are softened more rapidly than shallow slopes by diffusive processes. Diffusive obliteration times (for linear diffusion) scale as $D^2$. Therefore, a crater 5× the diameter of another will survive 25× as long, if diffusion is responsible for obliterating craters. This increases α by 2 (Eqn. 1). Therefore, the CSFD allows steady exhumation and/or mantling by dust, sand or ash (α increased by 1) to be distinguished from the diffusive alternative (Table 1). In practice, we find that most of our CSFDs are better fit by "α increased by 1" than by diffusive-obliteration (Figure 7a). Because the images we study have relatively good bedrock exposure and because we mask out large sand-covered areas within those images, we think that mantling is inferior to steady exhumation as an explanation for the CSFD. Diffusion may be insufficient to obliterate craters on some Mars sedimentary rock terrains: along the *Opportunity* traverse, craters that have been completely infilled by diffusion can still be recognized in HiRISE images (Fenton et al. 2015).

|  | *Crater obliteration* |  | *No crater obliteration* |
|---|---|---|---|
| *Power-law exponent of crater size frequency distribution, α.* | *Shallower than production function by +2* | *Shallower than production function by +1* | *Production function* |
| *'Kinematic' process* | Diffusion | Vertical advection (steady exhumation, serial retreat of many scarps, or steady burial) | Passive landscape *or* one-big-pulse resurfacing event |
| *Physical processes ('dynamics') corresponding to this kinematic process (bold: most likely for light-toned Mars sedimentary rocks)* | Seismic shaking<br>Slope creep<br>**Active-layer creep processes**<br>Small-meteorite gardening<br>Transport-limited fluvial processes<br>Any transport process that has an efficiency proportional to local slope. | **Wind-induced saltation abrasion** (more likely for bedrock exposures)<br><br>**Mantling by dust, sand or ash** (unlikely for bedrock exposures) | Any process that modifies, but does not obliterate craters (e.g. aeolian infilling; Fenton et al. 2015)<br>**Retreat of a tall, steep scarp**<br>Lava flooding |
| *Comments* | HiRISE anaglyph often shows diffusively softened craters that are not detectable in mono HiRISE | Distinguishing between exhumation and mantling requires careful inspection of high-resolution images. | Because we use $D>20$ m craters, steady exhumation rates ≲5 nm/yr will be misattributed as "production function" because >3 Gyr is required for such slow-eroding terrains to reach steady state. |
| *Examples* | Fassett & Thomson (2014). | This paper. | Michael (2013). |

**Table 1.** Interpreting the slope of a Martian crater size-frequency distribution on sedimentary rocks.

(2) *What specific mechanisms are responsible for crater obliteration?*
By themselves, CSFDs cannot fingerprint specific mechanisms of crater obliteration: instead, detailed site studies are needed. However, for low-latitude Mars sedimentary rocks where crater obliteration <100 Ma is substantial, the most likely agents of bedrock erosion are aeolian processes (e.g. Bridges et al. 2014). For steep scarps, either undermining by wind erosion, or mass wasting, may be the rate-limiting step for scarp retreat. Erosion by other processes is less likely; young fluvial channels are absent, and glaciers and young lava flows are uncommon. Now-vanished dust/ash/ice cover could alter $\alpha$ by shielding rocks from small craters but not large craters (Weiss & Head 2015). However, small craters formed in cover layers on Mars frequently form pedestal craters (e.g. Schon & Head 2012), which are easily identified. Pedestal craters are uncommon or absent near the sites we investigate in this paper.

(3) *How much erosion is needed to obliterate a crater?* The required vertical erosion to censor a crater from a crater count (obliteration) is unknown. The number used in this paper is $\phi$ = 50% of initial crater depth (where depth ≈ 0.2×$D$ for $D$<1 km; Watters et al. 2015). Erosion rates are inversely proportional to this parameter.

(4) *How does target strength affect erosion rate estimates?* Identical hypervelocity impacts form different-sized craters in targets that have different strength (Dundas et al. 2010). Thus, craters above a given size will form more swiftly in a weak target (e.g. regolith) than in a dry target. For sedimentary rocks on Mars, unconfined compressive strength >5 MPa is reported (Okubo 2007, Grindrod et al. 2010). This strength, if interpreted as a rock-mass strength (as is reasonable for erosion of unjointed bedrock), implies similar crater sizes to those formed in the lavas that are used to calibrate crater flux models (Hartmann 2005, Holsapple & Housen 2007), minimizing the importance of target strength.

(5) *Do exhumed craters significantly affect CSFDs?* Previously buried and now-exhumed $D$<0.1 km craters (Edgett & Malin 2002) are much less common than erosion-era small craters (Kite et al. 2013b). Possible causes include dilution by rapid paleo-sedimentation (Lewis & Aharonson 2014), disintegration of small impactors in past thicker atmospheres (Kite et al. 2014), and erosion during the era of net sedimentary rock emplacement (Sadler & Jerolmack 2007). Therefore exhumed craters are unlikely to corrupt our analysis.

(6) *Does steady exhumation discriminate between scarp retreat and vertical abrasion?* Steady exhumation is recognized by its effects on crater statistics gathered from an extended area. This necessarily averages over small-scale geologic processes. Specifically, if retreating scarps are numerous, randomly spaced, and less tall than the depth of the smallest crater in the sample, then the effect of scarp retreat on CSFDs is effectively indistinguishable from the effect of vertical abrasion on CSFDs. At Yellowknife Bay, the inferred retreating scarp is not tall enough to be distinguished from vertical abrasion using CSFDs (Farley et al. 2014). Both scenarios will average out (spatially and temporally) to "steady exhumation."

## 3. Workflow.

Our starting point is a list of crater diameters and a measure of the total area over which craters were counted. We assume that cratering is a Poisson process with fixed depth/diameter ratio 0.2 (Melosh 1989). We assume a crater flux model based on lunar counts and adapted to Mars, including an atmospheric-filtering correction (Hartmann 2005 as corrected by Michael 2013). (See §7 for a discussion of alternatives.) We set a minimum diameter $D_m$ based on visual inspection of the incompleteness turnoff (Figure S1). Next we carry out a power-law fit (Clauset et al. 2009):

$$\alpha_e = 1 + n \, [\Sigma_i \ln (D_i/D_m)]^{-1} \qquad (2)$$

where $\alpha_e$ is the estimate of $\alpha$, and which (for large $n$) has error well-approximated by

$$\sigma = (\alpha_e - 1)/N^{1/2} \qquad (3)$$

We then use the value of $\alpha_e$ to identify erosion processes by comparison to the crater-production function of Hartmann (2005) as corrected by Michael et al. (2013). The slope of the crater-production function steepens with increasing $D$, but the change in slope is much less than 1 in the range 22m < $D$ < 250m and so $\alpha_e$ can be used to identify erosion processes.

For sites which are adequately characterized by steady exhumation, we estimate erosion rate. For each crater-size bin, we normalize by count area to obtain the crater density, divide by the production-function to obtain the implied age, and divide the obliteration depth by the implied age to get the implied (binwise) erosion rate.

$$E_D \approx 0.2 \, D \, \phi \, / \, (N_D / f_D \, a) \qquad (4)$$

where the subscript $D$ denotes erosion rate at a given center-bin diameter, $N_D$ is the number of craters in the size bin, $f_D$ is the crater flux, and $a$ is the count area. The method does not take account of "bowl shrinkage" as the crater is ground down, which (for $\phi$ = 0.5 and a hemispheric crater) reduces diameters by <10%. The "≈" symbol refers to the fact that the median crater in a size bin is ~3% smaller than the bin-center (geometric center) diameter. For $\chi^2$ fitting of observed rates, we calculated the craters predicted in each bin for a wide sweep of obliteration rates:

$$N_D' = 0.2 \, D \, \phi \, a \, f_D \, / \, E^* \qquad (5)$$

where the prime denotes "predicted" and $E^*$ corresponds to trial obliteration rate. We obtained $\chi^2$ confidence intervals using standard methods (e.g. Wall & Jenkins 2012).

## 4. Example Dataset.

### 4.1. How the data were gathered.

We selected 18 HiRISE sedimentary rock images for analysis (Figure 3, Supplementary Table), focusing on images of light-toned layered deposits in Central Vallis Marineris and Gale (11 images), with the remainder selected from areas containing sedimentary rocks identified as Noachian/Early Hesperian materials of high potential for finding biological organic matter.

      Most studies involving crater counts rely on a single experienced analyst to identify craters. Robbins et al. (2014) compared lunar crater counts from 8 expert analysts to those of 1000s of non-specialist volunteers and found (i) expert-expert disagreement in crater density up to ±35% (ii) on average, non-specialists are able to identify craters as well as are expert analysts (see also Bugiolacchi et al. 2016). We took an intermediate approach by providing six analysts (University of Chicago undergraduates) with 2 hours of classroom training on martian impact crater morphology (with examples primarily drawn from HiRISE image data), followed by ~6 hours of hands-on training mapping impact craters on 2 HiRISE images using ArcMap and CraterTools (Kneissl et al. 2011). Following training, the analysts independently mapped craters in pre-selected areas of HiRISE images. We used map-projected image data from the HiRISE red channel as the basis for crater mapping. We typically selected either a band ⅓ the width of the image running from the top to the bottom of the image, or the entire image. Portions of the images containing dunes or other landforms of apparently unconsolidated material were masked out, leaving a count area of 546 km$^2$. The analysts (combined) made >5×10$^4$ crater identifications. Erosion rates were estimated for areas in HiRISE images in which craters were mapped by ≥3 analysts (Figure 4). For each such image, craters mapped by different analysts were aggregated using a clustering algorithm. The clustering algorithm tagged crater-pairs whose centers are separated by <50% of the maximum diameter, and whose radii differed by <50%, as being the same crater. These threshold percentages were obtained by trial-and-error inspection. A check loop ensured that "chained" craters (e.g. A-B is a pair, B-C is a pair, so A-B-C are a chain) were recorded as only a single, aggregated crater. Final agreed-upon craters were then defined by the mean center location and diameter of the clustered features (Figure 5). 1.3×10$^4$ craters were agreed by at least 2 analysts. The results define straight lines that are not well-fit by isochrons (Figure 6).

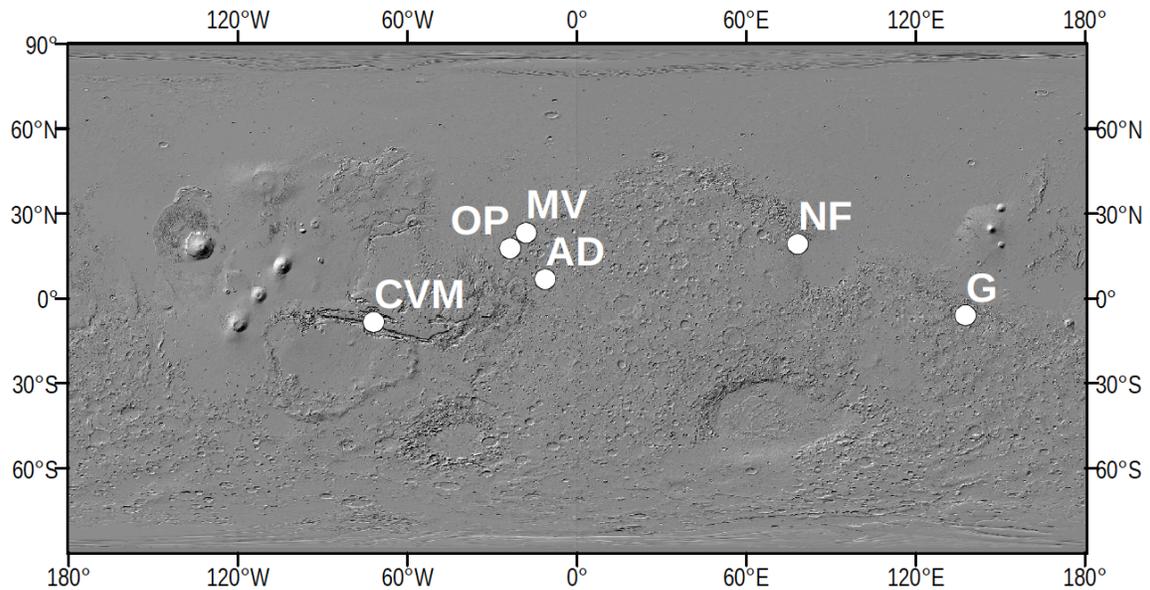

**Figure 3.** Map highlighting regions where craters were counted., CVM = Central Valles Marineris, OP = Oxia Planum, MV = Mawrth Vallis, AD = Aram Dorsum, NF = Nili Fossae (Nili Carbonates + SW of Jezero / "NE Syrtis"), G = Gale Crater.

To quantify expert-student divergence, an expert (D.P.M.) validated ∼⅓ of the counted area of 4 of the 18 HiRISE images. For these 4 scenes ($n$ = 308 craters with $D$ > 20m), we found a false-negative rate of 25% averaged over checked craters (worst case scene-average false negative rate 62%, best case 15%) and a false-positive rate of 9% averaged over all craters (worst case scene-averaged false-positive rate 43%, best case 6%) at the ≥2-agree level, where the worst case corresponds to a scene with only 7 craters. We chose to calculate obliteration rates based on the ≥2-agree case because it represents the smallest combined error rate relative to the expert reference. We multiply all of our counts by 1.34 to take account of the net error (false positive rate subtracted from false negative rate) from the aggregated checks. The checks showed a trend for students to undercount $D$∼50 m craters relative to both larger and smaller craters: we ignore this trend. We conclude that the student counts are accurate at the factor-of-2 level.

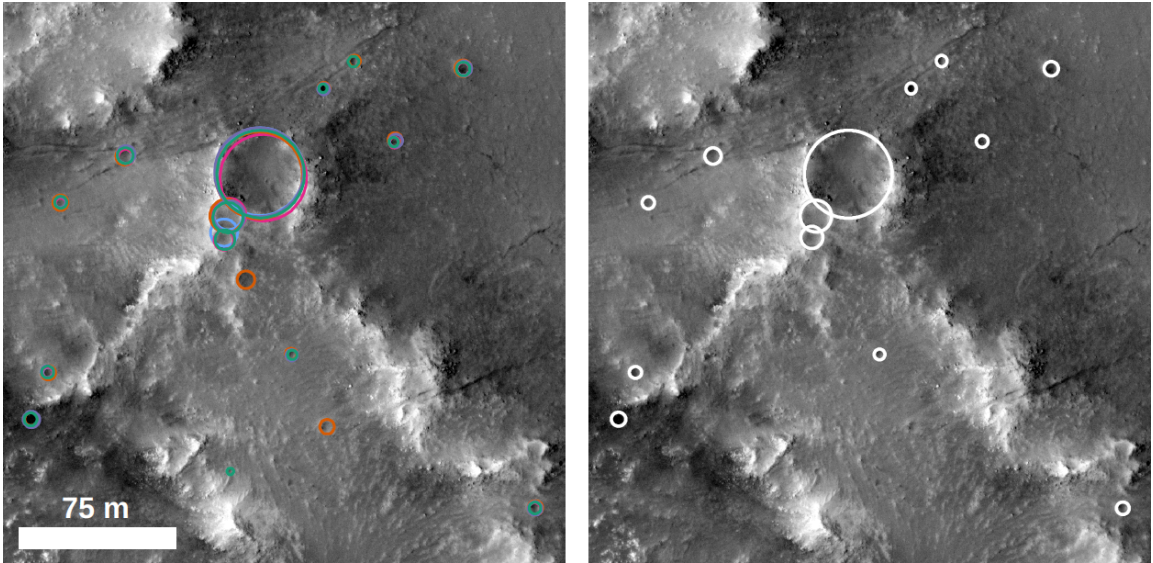

**Figure 4.** Example of how craters are aggregated. *Left:* Colors correspond to craters picked by an individual analyst. *Right*: Only agreed-upon craters have positions and diameters included in our analysis.

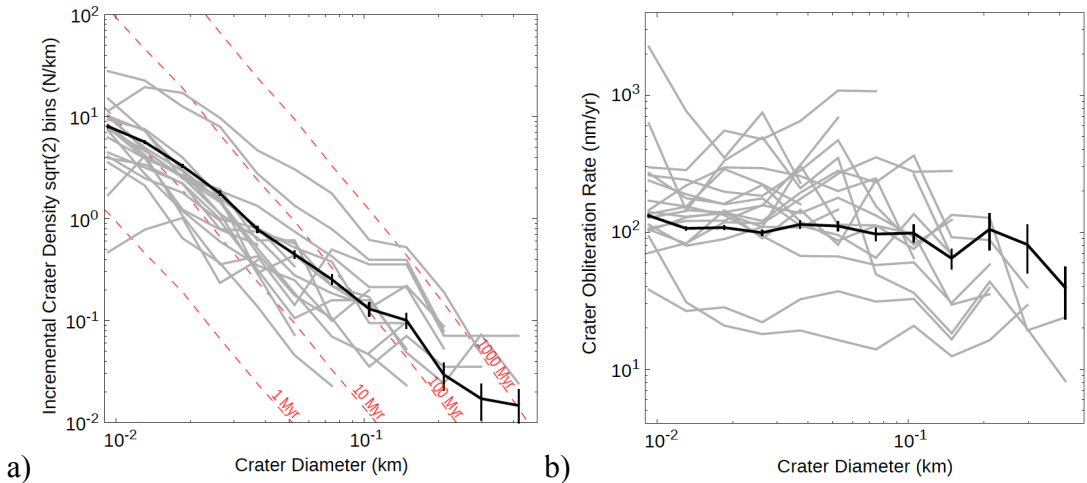

**Figure 5. (a)** Crater size-frequency distribution plot showing that our results are not well fit by isochrons. **(b)** The counts from (a), converted using Eqn. 4 to erosion rates. In aggregate, the results are well-fit by steady exhumation at ~100 nm/yr. Gray lines are data from individual HiRISE images and bold black lines are the average across all images. Although the data are from different geologic units, crater-obliteration rates fall in the range 10-1000 nm/yr.

## 4.2. Analysis of combined dataset.

The hypothesis of steady exhumation is supported by the combined dataset. For $D$<10m, we saw a sharp inflection in the CSFD that we attribute to survey incompleteness (Figure S1). To be conservative, we discard $D$<22m craters when fitting. For diameters 22-250m, the best-fit power law slope is -1.87, which is close to the -1.9 expected for steady exhumation (Figure 7a). The formal statistical errors for the power law slopes (±0.02 for data, ±0.05 for model) are likely smaller than the real errors. However, the data in Figure 5a are not fit by isochrons (dashed red lines) and are also not fit by diffusion (which would lead to an even shallower CSFD than observed). The best-fit exhumation rate is 102±7 nm/yr. Evidence for steady exhumation in the combined dataset (Figure 7b) is present over at least a decade in crater diameters (20m – 200m), corresponding to times 20-200 Myr in the past.

## 4.3. Regional variations.

The data show important regional variations from the global averages discussed above. HiRISE images from the same geographic region usually show similar values of $\alpha$ and $E$ (Figure 6). For the light-toned layered deposits (Central Valles Marineris and Gale), as well as for Oxia Planum, $\alpha$ is well-explained by steady exhumation. For these sites, the CSFD cannot be reproduced by a spatial mixture of patches where craters are obliterated by diffusive processes, and patches where the CSFD is the same as the production function (zero-obliteration, unmodified). Such a spatial mixture model can be tuned to fit any value of α. However, a two-parameter spatial mixture of diffusive degradation and single-resurfacing predicts concave-up curvature in the bin-by-bin plots on a log scale. Curvature is not observed for the light-toned layered deposits, nor for Oxia Planum. Because α is well-explained by steady exhumation, and the alternative spatial-mixture model fails to explain the CSFD, the CSFD for these sites is suggestive of steady exhumation.

Among sites with α consistent with steady exhumation, global exhumation-rate variations at the image level are modest (Figure 7b). Most confidence intervals overlap the range $E$ = 100-1000 nm/yr (since 10-100 Mya). This modest variation is surprising if saltation abrasion by sand is responsible for exhumation, because sand is found mostly at low elevations. The modest variations could be because the abrasive particles responsible for saltation abrasion are sourced locally, or it could indicate a different process (for example, deflation of weathered fragments) is the rate-limiting step for bedrock erosion, or it could be a coincidence.

Faster erosion is indicated for 2 images (PSP_006190_1725 and PSP_003896_1740) from Candor Chasma in Central Valles Marineris (Figure 7b).

The SW Melas data show very low crater density at topographic elevations that were repeatedly flooded by lakewaters (Metz et al. 2009, Williams & Weitz 2014). This is offset in our image averages by relatively high crater density near the margins of the paleolake, so that the image-averaged erosion rate is not unusual (Figure 6). Because the parts of SW Melas that were underwater for the longest time are the zones of greatest astrobiological interest (Metz et al. 2009), we used HRSC DTM H2138_0000 (50 m/pixel) to clip out terrain lying below the -2250m contour. The -2250m contour corresponds to the lowest candidate lake level discussed by Williams & Weitz (2014). For this low-lying terrain, we did a separate erosion rate fit. We found α = 2.3±0.8 (2σ), which is consistent both with a single resurfacing event and with steady exhumation. For steady exhumation, the fit is $E$ = 530 nm/yr, (95% confidence interval 320-870 nm/yr, $n$ =44 craters, combined area = 19 km$^2$).

Oxia Planum and Aram Dorsum show a low erosion rate relative to the the other sites for which we fit $E$: 10-30 nm/yr. Although the crater density is noticeably variable between geologic units, all show a high density.

Steady exhumation is not a sufficient explanation of CSFDs at "NE Syrtis" or Mawrth. The CSFDs show a power-law slope that is shallower than expected for steady exhumation. Here, another process is required to rapidly obliterate small craters. Possibilities include aeolian infilling (by small patches of bedforms not included in the sand mask). Therefore we do not interpret the estimated exhumation rates for "NE Syrtis" and Mawrth to be realistic.

At the Nili Carbonates site, the steady exhumation hypothesis is not rejected at the 95% level (Figure 7a). However, we suspect that diffusive processes, and obscuration by small bedform patches not included in the sand mask, are a major contributor to crater nondetection.

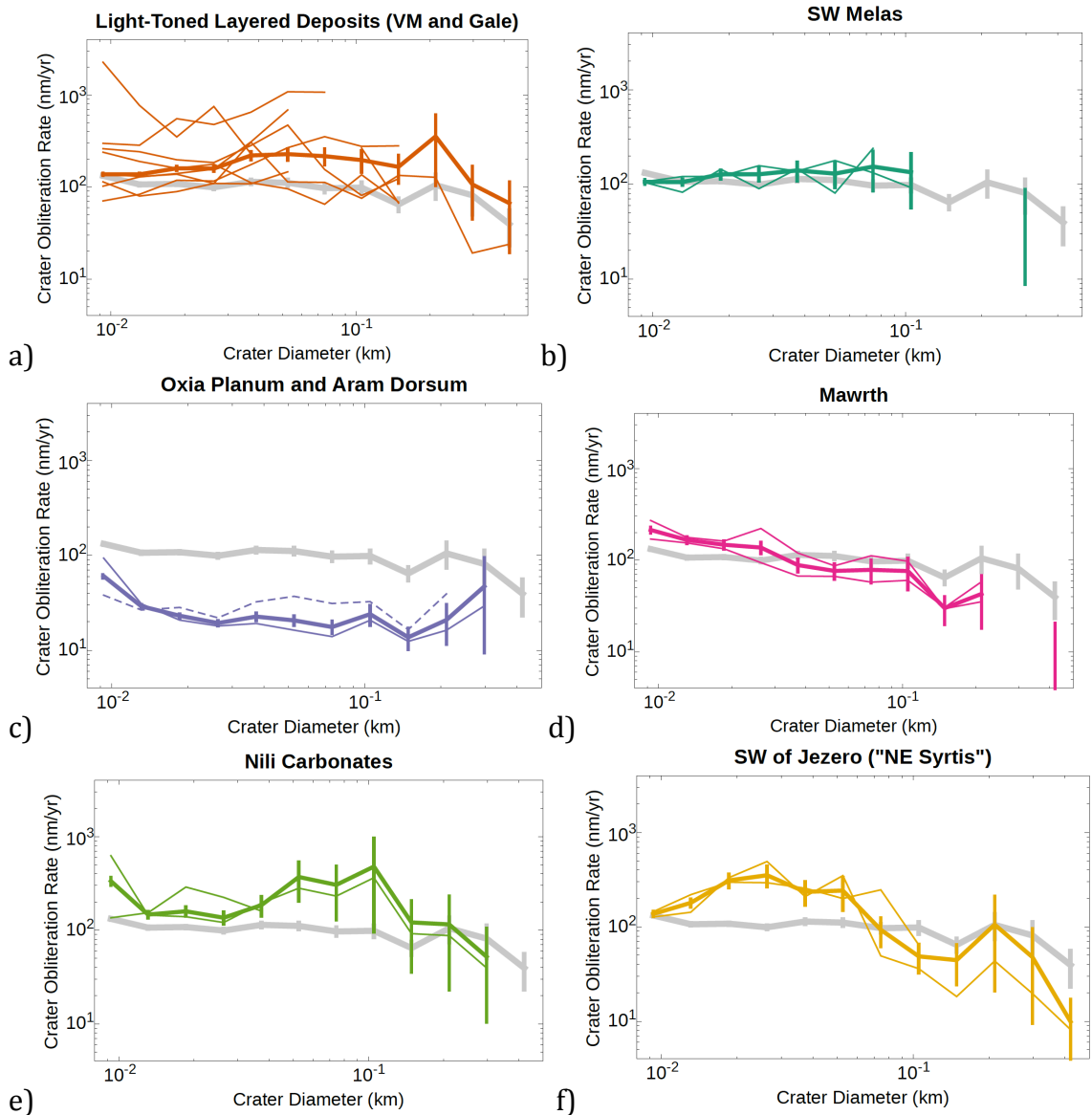

**Figure 6.** Crater-obliteration rate for different geologic settings. Thin lines show data collected from individual HiRISE images. Thick lines aggregate data from multiple images for a single geologic setting. (For size bins where some images show no craters, the thick lines can plot above all thin lines due to count-area.) 1-σ error bars shown at intervals of $2^{1/2}$ in *D*. Thick gray line shows crater-obliteration rate for all data. **(a)** Light-toned layered deposits in Valles Marineris and Gale ($3.1 \times 10^2$ km$^2$). **(b)** SW Melas paleolake deposits ($4 \times 10^1$ km$^2$) **(c)** Oxia Planum (thin solid line) and Aram Dorsum (thin dashed line) (combined area $5 \times 10^1$ km$^2$) **(d)** Mawrth ($6 \times 10^1$ km$^2$) **(e)** Nili Carbonates ($5 \times 10^1$ km$^2$) **(f)** SW of Jezero ("NE Syrtis") ($5 \times 10^1$ km$^2$).

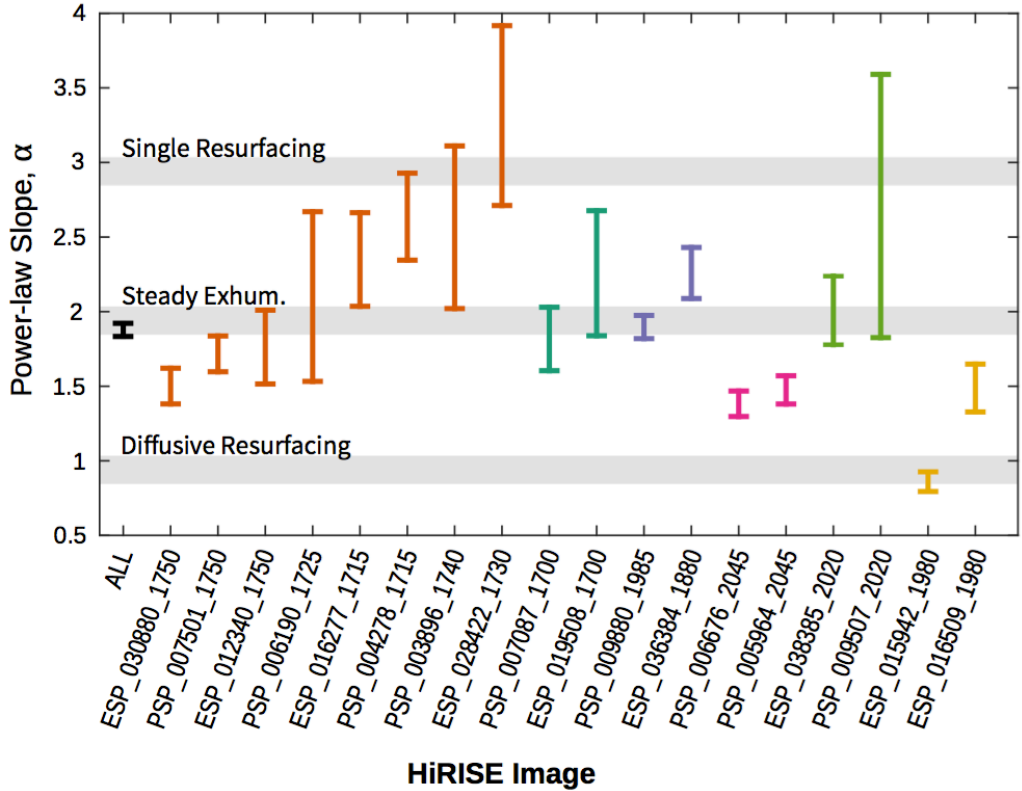

(a)

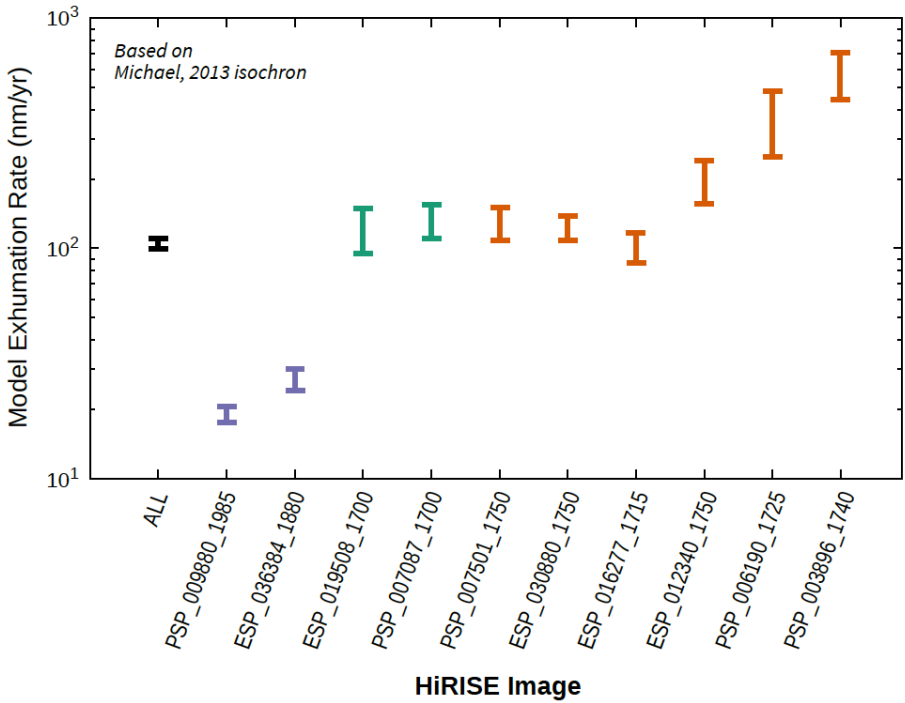

(b)

**Figure 7.** Site-by-site listing of erosion parameters. Error bars bound 95% confidence intervals. (a) Size-frequency power-law exponent α. (b) Exhumation rate for the subset of sites that have a value of α and/or a geologic expression that we interpret as being consistent with steady exhumation. Assumes 50% obliteration depth fraction. Colors correspond to those used in Figure 6.

## 5. Application to landscape evolution and the global dust cycle.

### 5.1. Landscape evolution.
Now we return to the landscape evolution question raised in §1. Light-toned layered sedimentary rocks on Mars often form mounds with 2-8 km of relief. Does recent erosion conserve, enhance, or reduce this relief?

Our measurements show fast erosion rates near the summits of major sedimentary rock mounds (e.g. Ceti Mensa, Juventae Mensa). These sites often show large exposures of relatively-young light-toned regularly-bedded rocks, termed "rhythmite" by Grotzinger & Milliken (2012) (e.g., PSP_003896_1740 and PSP_006190_1725). Images within the mounds, but below the rhythmite, show lower steady-exhumation rates (e.g., ESP_012340_1750). By contrast, images within moats show higher crater densities, with $\alpha$ corresponding not to pure steady exhumation but rather to a mix of processes including a component of production function (e.g., PSP_004278_1715). The contrast between faster erosion of rhythmite and slower erosion of older rocks would be even greater if we modeled atmospheric filtering as a function of elevation. The contrast between rapid erosion on mountain summits and slow (or no) erosion in moats suggest that mound relief is currently being reduced by wind erosion.

In addition to the spatial gradients in erosion rates discussed above, our data suggest changes in wind-erosion rates with time. If our erosion rates for Gale's mound (Mt. Sharp) correspond to the Amazonian average, then only modest (~0.5 km) exhumation of Mt. Sharp has occurred since 3 Gya. Because Mt. Sharp records $\gg 0.5$ km of wind erosion, this indicates a higher rate of wind erosion in the past. Therefore, the sculpting of Mt. Sharp cannot be explained by present-day rates and processes.

### 5.2. Dust cycle.
If craters are destroyed purely by landscape lowering, then multiplying our crater obliteration rates by the area of light-toned layered sedimentary rock outcrops on Mars (~$2 \times 10^6$ km$^2$) yields a production rate of fine-grained sediment of $10^{-4}$ km$^3$/yr or a ~4 m global equivalent layer if erosion was sustained over 3 Gyr. This is an underestimate, because we do not attempt to constrain erosion rates on the steep scarps bounding the Valles Marineris interior layered deposits and these erosion rates could be very high. The fine-grained sediment would be mostly dust due to grain attrition (Jerolmack & Brzinski 2010, Cornwall et al. 2015). In reality Mars is not covered by a ~4m global equivalent layer of dust, so the volume estimate suggests the existence of dust sinks. Specific candidate sinks include (from most to least voluminous) rhythmite including the Medusae Fossae Formation (Tanaka 2000), Planum Boreum's Basal Unit (Byrne & Murray 2002), duststone on the Tharsis plateau (Bridges et al. 2010), dust deposits in Arabia Terra (Mangold 2009), and dust-rich layers trapped between young lavas in Amazonis (Morgan et al. 2015). If Mars dust today is volumetrically trivial compared to all dust released since 3 Ga, then the presently circulating dust on Mars should have been mobilized <<1 Ga. This hypothesis might be tested comparing dust composition to sedimentary rock composition. Mars dust is described by Goetz (2005) and Pike et al. (2011) as being dominantly mafic in composition. This could imply physical separation of the primary and secondary minerals.

A plausible sink for Mars dust is rhythmite. That would imply dust-to-dust recycling (aeolian cannibalism) (Kerber & Head 2012), as seen on Earth (Licht et al. 2016). If sediment produced by erosion of layered deposits is re-incorporated into rhythmite in <<3 Ga, then the reduction in mound relief suggested for present mounds need not be representative of the long-term average.

**6. Application to preservation of complex organic matter.**

**6.1. Background.**
We now return to the organic-matter preservation question raised in §1. Complex organic matter is much more biologically diagnostic than simple organic matter, but also much more vulnerable to radiolysis by galactic cosmic radiation (Mustard et al. 2013, Pavlov et al. 2012). Simple organic materials (e.g. chlorobenzene, glycine) can be produced by both biotic and abiotic processes. Simple abiotic organic matter is found at percent level in several meteorite classes, and meteorites would have dusted Mars at a high rate early in Mars history – "cosmic pollution" that complicates the search for past life on Mars (Summons et al. 2011). If simple organic matter exists in Mars mudstones (Freissinet et al. 2015), it would be difficult to diagnose biological versus nonbiological origin. By contrast, complex organic molecules can be unique biomarkers in ancient sediments (Mustard et al. 2013). That is because of their ability to preserve unique identifiers of biology such as the use of repeating subunits. Unfortunately, potential biomarkers are much more vulnerable to radiolysis than is simple organic matter (Pavlov et al. 2012). Larger-molecular-weight amino acids are more vulnerable to radiolysis (Kminek & Bada 2006), which maps to a much smaller survival fraction because of the exponential-decay nature of radiolysis. Therefore ancient complex organic matter is best sought in rocks that have received a minimal radiation dose.

**6.2. Method.**
The cumulative radiation dose experienced by a rock depends on its erosion/exhumation history: deep burial is an effective radiation shield. Assuming steady exhumation, the cumulative radiation dose $R$ is

$$R \approx \int a e^{-bz} dt = \int_0^{z_0/E} a \exp(z_0 b - z_0 b t E) dt = -\frac{a \exp(-z_0 b)}{bE}$$
(6)

where $a$ and $b$ are fit to the depth-dependent calculations of Hassler et al. (2014), and $z_0$ is a negligible-GCR depth (we use 100m). The survival fraction of organic matter $\Omega$ is then given by

$\Omega = \exp(-k_m R)$ (7)

where $k_m$ is radiolysis constant (Gy$^{-1}$), and $m$ is molecular mass (Da). We ignore inherited radiation damage (in effect, we assume swift burial or a >100 mbar deposition-era atmosphere).

### 6.3. Results.

We find that in the most optimistic case ($k_m$ from Kminek & Bada 2006; thin dashed lines in Figure 8), 100 nm/yr exhumation gives a GCR dose that would reduce complex organic matter abundance 2-fold. Experiments using amino acids within geologic analog materials find much worse preservation potential than for purified amino acids, especially when H is present as is certain for Mars soil (Pavlov et al. 2016). More realistic decay constants (e.g. $k_m$ from Pavlov et al. 2016, thick line) would worsen this to a ≥10-fold reduction.

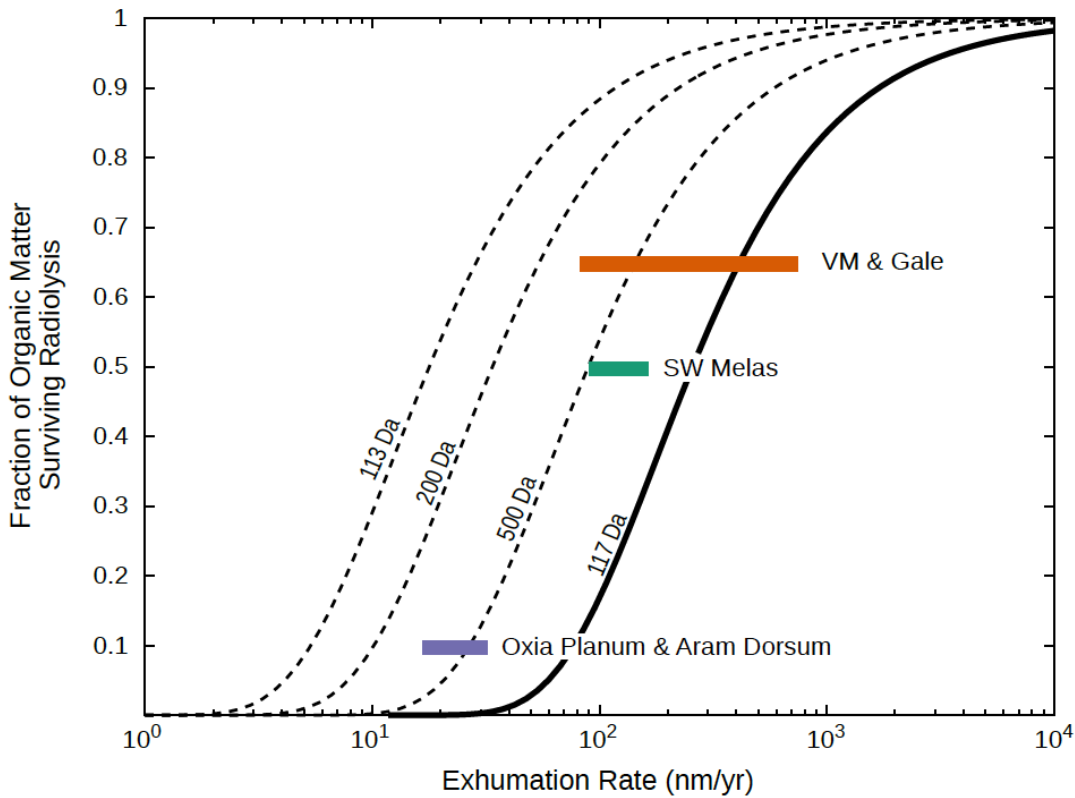

**Figure 8.** Radiolysis survival chart for organic matter currently found at 3 cm depth. The thin dashed lines show organic-matter preservation for atomic mass 113, 200, and 500 Da, based on experimental data for pure amino acids (Kminek & Bada 2006). The thick black line shows an upper limit on preservation of L-isovaline (117 Da) in SiO$_2$ (Pavlov et al. 2016). Horizontal bars show the ranges of exhumation rate estimated for 3 different geologic settings.

Oxia Planum is the worst place (in terms of radiolysis) to test for the past presence of biogenic organic matter on Mars.

    The results of our radiolysis modeling indicate that Oxia Planum and Aram Dorsum could preserve only a small percentage of complex organic matter (Figure

8) based on their calculated exhumation rates. This may affect the suitability of these areas as proposed landing sites for future robotic missions to search for preserved organic matter. However, larger fractions of organic matter are expected to be preserved below the radiolytically-processed layer (> 2 m depth), even in areas where erosion rates are relatively low. Similarly, larger fractions of organic matter may be preserved in areas where erosion rates are locally higher, such as on retreating scarps.

## 7. Discussion.

### 7.1. Comparison to other datasets and Earth analogs.

Our 100 nm/yr erosion estimate is in accord with independent methods. Horizontal retreat rates from landslide-molds (Grindrod & Warner 2014), converted to vertical erosion, yield 300-800 nm/yr vertical abrasion. This is at the high end of our measurements. That is understandable because steep slopes retreat more rapidly and because the landslide-mold approach only works in regions of rapid erosion. Our erosion rates exceed those of Golombek et al. (2014), but these were obtained for flat landscapes. We suspect that (as on Earth; Larsen et al. 2014), steep terrains contribute most of the eroded flux. Steady-exhumation fits are consistent with the erosion rate due to scarp retreat inferred from cosmogenic isotopes (Farley et al. 2014).

At Mars landing sites, textural evidence for rock erosion by saltation abrasion is ubiquitous (Bridges et al. 2014). On Earth, aeolian deflation can be important for basin exhumation in dry settings (Rohrmann et al. 2013). In Antarctica, saltation-abrasion rates can reach ~30000 nm/yr for basalt and sandstone (Malin 1985). Scaling of these measurements to Mars indicates landscape-lowering rates of 900-9000 nm/yr (Bridges et al. 2012). These calculations are for sand fluxes within a dunefield, so likely overstate the long-term abrasion rate. Nevertheless, at the rates we infer, saltation abrasion is a reasonable explanation for steady exhumation.

For softer materials (potentially including rhythmite), it is possible that decomposition of cementing minerals, thermal cycling, or removal of loose particles by the wind, may be more important than saltation abrasion in setting the pace of erosion.

### 7.2. Which crater-production function to use?

In this paper, the starting point for our workflow is crater-flux models based on lunar counts and adapted to Mars (Hartmann 2005 as corrected by Michael 2013), rather than empirical estimates of crater-flux based on observations from Mars orbit of craters that formed over the last ~10 years (Daubar et al. 2013). That is because the observed present-day crater flux is spatially nonrandom, even after correcting for monitoring efficiency (Daubar et al. 2014). Because the true flux is spatially almost random (Le Feuvre & Wieczorek 2008), the documented nonuniformity shows that a spatially varying target property masks a subset of currently-forming craters from detection. Therefore the Daubar et al. (2013) flux is a lower limit, whereas the Hartmann (2005) flux is a best estimate. Given that spatially-varying masking occurs, it would be expected to mask a greater fraction of small craters than large craters. Therefore, size-dependent masking may explain some (or all) of the shallower slope of the Daubar et al. (2013) flux relative to the Hartmann (2005) flux (Williams et al. 2014, Daubar et al. 2016). Another possible contributor to the difference in slope may be that the Hartmann (2005) flux includes distant secondaries, whereas the Daubar et al. (2013) counts likely do not. However, Williams et al. (2014) show that primaries alone can reproduce the Hartmann (2005) flux, suggesting that the role of secondaries is minor. Perhaps more importantly, variations in flux are expected due to changes in Mars' orbit (JeongAhn & Malhotra 2015). Therefore, the true spacecraft-era crater-flux may not be representative of the flux averaged over many orbital cycles that is relevant for our calculations.

It is conceivable that (contrary to the arguments given above) the numbers reported by Daubar et al. (2013) are in fact representative of crater flux over the last ~10 Myr. If so, our disfavoring of the landing sites listed in §6 is sharpened, and our steady exhumation hypothesis must be discarded in favor of a hypothesis of serial retreat of scarps >20m tall.

Like all Mars crater-chronology work, ours suffers from being calibrated to lunar data. However, modeling of masking processes and of variations in flux over orbital timescales (JeongAhn & Malhotra 2015), continued monitoring of the modern impact flux, and future direct dating of surfaces on Mars, will all abrade these uncertainties.

**7.3. Outlook.**

The small-crater record is potentially a powerful proxy for erosion and deposition over the last ~1 Gyr on Mars. To fully exploit this proxy requires progress on (1) systematics, (2) process validation, (3) mapping the spatial variability of erosion rates and interpreting the results.

(1) The main systematics are the true $10^7$-yr average flux of bolides arriving at Mars (JeongAhn & Malhotra 2015) and inter-analyst variability. Uncertainties in the effect of the present-day atmosphere on bolides are relatively minor (Williams et al. 2014). However, it is possible that the average atmospheric pressure over the last $10^8$ yr was higher than today, and this could greatly reduce our erosion rates. On the other hand, if the rock-mass strength of the sedimentary rocks is <5 MPa, then the craters correspond to smaller (and thus more common) impactors, and and this would increase our estimated erosion rates. As one example of inter-analyst variability, we found that students sometimes underestimated crater diameters. To assess this, we recalculated obliteration rates based on ≥2-agreed craters after increasing the diameters of a random sample of 25% of craters by 50%. The recalculation decreased obliteration rates by ~30% (which would worsen organic-matter preservation). Inter-analyst variability may be mitigated by increasing the number of analysts.

(2) To verify the process hypothesis (steady exhumation) advanced in this paper, co-mapping of crater textures and CSFDs would help. How does crater morphology change as the crater power-law slope changes? What is the crater size-frequency distribution when both diffusive obliteration and landscape-lowering are active? How do exhumation-rate inferences change when stochastic models (e.g. Richardson 2009) that capture crater-diameter change during erosion are included? Combining the $\kappa = 10^{-6}$ m$^2$ yr$^{-1}$ Mars crater-degradation diffusivity reported by Golombek et al. (2014) with our $E = 10^{-7}$ m yr$^{-1}$ steady exhumation rate suggests a length scale of $\kappa/E = 10$m at which the two processes balance.

(3) In many cases the crater density appears to vary in an obviously nonrandom way at scales that the image averages shown in Figure 6 do not capture. This raises unanswered questions. What causes these variations? To what extent does erosion rate vary at 1 km scale? At 10 km scale? Is lithology, or the surrounding terrain, more important? How well does crater density correlate with independent measurements of erosion resistance (e.g. topographic protrusion; Becerra et al. 2016)? Our dataset covers small patches of Mars, allowing tentative hypotheses about the controls of erosion. Testing these hypotheses would be prohibitively expensive if the approach taken is to greatly increase the number of HiRISE counts (unless volunteers can be involved; Bugiolacchi et al. 2016). One logical next step is a CTX-based study covering large patches of fast-eroding Mars terrain. Such a study might use geologic gradients as natural experiments to test for the relative importance of the main geologic factors affecting erosion rate: lithologic variations in erosion susceptibility, paleoatmospheric variations, changes in the erosion rate with time, terrain effects (Kite et al. 2013a), and variations in the supply of abrasive sand. In our images, crater density within-images is highly variable (see also Warner et al. 2015).

## 8. Conclusions.

- We provide a workflow for using crater counts to constrain crater-obliteration rates on Mars, making use of small craters whose size-frequency distribution does not follow isochrons.

- Using the Hartmann (2005) crater flux, the crater-obliteration rate of light-toned layered sedimentary rocks is $\sim 10^2$ nm/year. Based on crater morphology and bedrock exposure frequency, we interpret this as an exhumation rate of $\sim 10^2$ nm/year.

- Our results suggest that the relief of Mars' major sedimentary rock mounds is currently being reduced.

- The exhumation rate at the paleolake deposits in SW Melas Chasma is relatively high. Therefore radiolysis is less of a threat to the preservation of ancient, complex organic matter at these paleolake deposits than at the other sites investigated. The exhumation rate at Oxia Planum and Aram Dorsum is relatively low. Therefore radiolysis is more of a threat to the preservation of ancient, complex organic matter at Oxia Planum and Aram Dorsum than at the other sites investigated. Assuming steady-state exhumation and a pure amino-acid target, we provide quantitative estimates of best-case preservation for complex organic matter.

**Acknowledgements.** We gratefully acknowledge the undergraduate research assistants who counted craters for this project: Daniel Eaton, Julian Marohnic, William Misener, Emily Thompson, Edward Warden and Chuan Yin. We thank Jean-Pierre Williams for help and advice on atmospheric-filtering corrections. We also thank Cathy Quantin, Eliot Sefton-Nash, Alex Pavlov, Ingrid Daubar, Jasper Kok, Tim Michaels, Mark Allen, Pamela Gay, and Nathan Bridges, for discussions and unselfish sharing of data. This work was financially supported by the U.S. taxpayer (NASA grant NNX15AH98G).

**Supplementary Table 1.** List of images used.

| Geologic classification | Geographic region | HiRISE image number |
|---|---|---|
| Light-toned layered deposits | Gale | ESP_030880_1750 |
| Light-toned layered deposits | Gale | PSP_007501_1750 |
| Light-toned layered deposits | Gale | ESP_012340_1750 |
| Light-toned layered deposits | Valles Marineris (VM) | PSP_006190_1725 |
| Light-toned layered deposits | Valles Marineris (VM) | ESP_016277_1715 |
| Light-toned layered deposits | Valles Marineris (VM) | PSP_004278_1715 |
| Light-toned layered deposits | Valles Marineris (VM) | PSP_003896_1740 |
| Light-toned layered deposits | Valles Marineris (VM) | ESP_028422_1730 |
| Paleolake and associated deposits | SW Melas Chasma | PSP_007087_1700 |
| Paleolake and associated deposits | SW Melas Chasma | ESP_019508_1700 |
| Sedimentary deposits | Oxia Planum | PSP_009880_1985 |
| Sedimentary deposits | Aram Dorsum | ESP_036384_1880 |
| Al/Mg/Fe-phyllosilicate plains | Mawrth | PSP_006676_2045 |
| Al/Mg/Fe-phyllosilicate plains | Mawrth | PSP_005964_2045 |
| Carbonate-bearing plains | Nili Carbonates | ESP_038385_2020 |
| Carbonate-bearing plains | Nili Carbonates | PSP_009507_2020 |
| Diverse materials including sedimentary rocks | SW of Jezero ("NE Syrtis") | ESP_015942_1980 |
| Diverse materials including sedimentary rocks | SW of Jezero ("NE Syrtis") | ESP_016509_1980 |

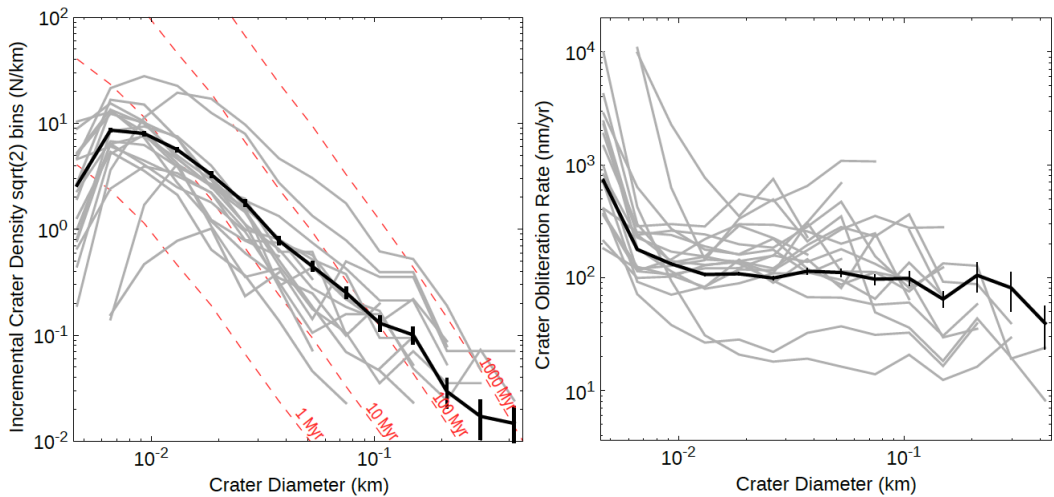

**Figure S1.** As for Figure 5 in the main text, but extended to show the reduction in crater abundance at small diameter due to survey incompleteness. The individual images show an incompleteness rollover at diameters ranging from 5m to 20m; to be conservative, we exclude all craters smaller than 20m. *Left panel:* Crater size-frequency distribution plot showing that our results are not well fit by isochrons. *Right panel:* The counts from the left panel, converted using Eqn. 4 to erosion rates. In aggregate, the results are well-fit by steady exhumation at ~100 nm/yr. Gray lines are data from individual HiRISE images and bold black lines are the average across all images.

**Supplementary Table 2.** Raw count data (for 2-agree craters), uncorrected for analyst undercounting. Incremental counts in √2 bins.

| Image orbit | Bin minimum (m) | | | | | | | | | | | | | | | Count area (km) |
|---|---|---|---|---|---|---|---|---|---|---|---|---|---|---|---|---|
| | 3.9 | 5.5 | 7.8 | 11 | 15.6 | 22.1 | 31.3 | 44.2 | 62.5 | 88.4 | 125 | 177 | 250 | 354 | 500 | |
| PSP_003896 | 56 | 234 | 155 | 92 | 28 | 16 | 6 | 2 | 1 | 0 | 1 | 0 | 0 | 0 | 0 | |
| PSP_004278 | 118 | 717 | 646 | 310 | 129 | 65 | 22 | 8 | 3 | 2 | 1 | 0 | 0 | 0 | 0 | 58 |
| PSP_005964 | 43 | 127 | 117 | 73 | 50 | 35 | 25 | 14 | 8 | 4 | 4 | 1 | 0 | 1 | 0 | 58 |
| PSP_006190 | 0 | 2 | 6 | 10 | 13 | 3 | 5 | 0 | 2 | 0 | 0 | 0 | 0 | 0 | 0 | 25 |
| PSP_006676 | 15 | 55 | 89 | 77 | 50 | 18 | 17 | 13 | 5 | 3 | 5 | 2 | 0 | 1 | 0 | 17 |
| PSP_007087 | 114 | 290 | 214 | 107 | 63 | 24 | 14 | 6 | 4 | 3 | 0 | 0 | 1 | 0 | 0 | 31 |
| PSP_007501 | 32 | 263 | 311 | 190 | 106 | 66 | 12 | 18 | 9 | 7 | 2 | 1 | 3 | 1 | 0 | 29 |
| PSP_009507 | 4 | 46 | 70 | 35 | 11 | 7 | 5 | 0 | 0 | 0 | 0 | 0 | 0 | 0 | 0 | 55 |
| PSP_009880 | 4 | 76 | 238 | 408 | 358 | 204 | 98 | 64 | 37 | 13 | 11 | 4 | 1 | 0 | 0 | 12 |
| ESP_012340 | 19 | 121 | 77 | 47 | 34 | 18 | 6 | 2 | 3 | 3 | 1 | 0 | 0 | 0 | 0 | 28 |
| ESP_015942 | 146 | 180 | 117 | 59 | 15 | 5 | 6 | 2 | 7 | 5 | 5 | 1 | 1 | 1 | 0 | 25 |
| ESP_016277 | 17 | 174 | 192 | 155 | 82 | 33 | 17 | 7 | 0 | 1 | 2 | 0 | 0 | 0 | 0 | 19 |
| ESP_016509 | 39 | 268 | 147 | 55 | 24 | 12 | 7 | 5 | 2 | 4 | 0 | 0 | 0 | 1 | 0 | 28 |
| ESP_019508 | 50 | 120 | 99 | 71 | 24 | 19 | 6 | 6 | 1 | 0 | 0 | 0 | 0 | 0 | 0 | 27 |
| ESP_028422 | 246 | 426 | 287 | 129 | 70 | 31 | 8 | 2 | 0 | 0 | 0 | 1 | 0 | 0 | 0 | 13 |
| ESP_030880 | 97 | 128 | 94 | 67 | 47 | 21 | 17 | 11 | 8 | 2 | 2 | 0 | 0 | 1 | 0 | 37 |
| ESP_036384 | 60 | 272 | 353 | 286 | 159 | 101 | 35 | 17 | 10 | 5 | 5 | 1 | 0 | 0 | 0 | 28 |
| ESP_038385 | 0 | 4 | 48 | 117 | 72 | 41 | 13 | 5 | 3 | 1 | 2 | 1 | 1 | 0 | 0 | 17 |
| TOTAL | 1060 | 3503 | 3260 | 2288 | 1335 | 719 | 319 | 182 | 103 | 53 | 41 | 12 | 7 | 6 | 0 | 38 |